\documentclass[10pt,letterpaper]{article}
\usepackage{opex3}

\usepackage{graphicx,amsmath,amssymb}
\begin{document}

\title{
Design of resonant microcavities: application to optical gyroscopes 
}

\author{Satoshi Sunada and Takahisa Harayama}

\address{
Department of Nonlinear Science, ATR Wave Engineering Laboratories,
2-2-2 Hikaridai Seika-cho Soraku-gun, Kyoto, 619-0228 Japan\\
}

\email{sunada@atr.jp} 



\begin{abstract}
We study theoretically and numerically the effect of rotation on
 resonant frequencies of microcavities in a rotating frame of reference.
Cavity rotation causes the shifts of the resonant frequencies
 proportional to the rotation rate
 if it is larger than a certain value.
Below the value, a region of rotation rate exists 
where there is no resulting the frequency shifts proportional to the rotation rate.
We show that designing cavity symmetry as $C_{nv}$ ($n\ge 3$) can 
eliminate this region.
\end{abstract}

\ocis
{
(140.4780) Optical resonators, 
(140.3410) Laser resonators, 
(140.3370) Laser gyroscopes,
(350.5720) Relativity.
} 


%
%
%
\section{Introduction}
The propagation of light fields and the resonances of optical
cavities in rotating systems have been experimentally and 
theoretically studied as one of the most basic and interesting problems
of electromagnetics in arbitrary accelerated systems
\cite{Post}.
Many of these investigations have concentrated 
on rotating ring
interferometers and rotating ring cavities
because such devices can be applied to optical rotation velocity sensors
\cite{Post,Crow,Aronowitz,Lamb2,Kuriyagawa,Arditty}.

One of the first suggestions to optically detect rotation rate was 
given by Sagnac in 1913 \cite{Post}. 
The basic idea is to use the difference in one round-trip time between 
 clockwise (CW-) and counter-clockwise (CCW-) propagating waves due to
rotation in a ring interferometer.
The time difference was experimentally observed as phase difference.
Today the measurement of 
 phase difference in ring interferometers has become very precise 
due to optical fiber technology.  

On the other hand, rotation also affects the resonances of optical
cavities (modal structures).
The degenerate resonant frequency of a non-rotating ring cavity
was split into a doublet by rotation, and the fact was confirmed by the
experiments of Macek and Davis \cite{Macek}. 
The effect of the splitting has becomes the basic principle of the
operation of ring laser gyroscopes, which are 
now used in airplanes, rockets, and ships etc. 
for autonomous navigation because
they are the most precise sensors among any other types of
gyroscopes.

So far, the effect of rotation on resonances (i.e., frequency splitting
induced by rotation) has been explained as that 
which originated in the time difference 
between counter-propagating lights due to rotation. 
This fact has been derived by 
a kinematical approach (ray-optics) 
\cite{Post,Crow,Aronowitz,Lamb2,Kuriyagawa} or by a wave dynamical
approach based on the electromagnetic equations of a naturally covariant
form (or general theory of relativity) \cite{Post,Anderson,Landau,Heer}. 
For such theoretical approaches, 
the following assumptions are explicitly or implicitly used in their
derivations:
(i) The light field one-dimensionally propagates in slender waveguides
such as optical fibers or ring cavities composed of more than three
mirrors; 
(ii) The wavelength of the light is much shorter than the sizes of the
cavities, i.e., The geometrical optics approximation is valid enough.
The use of assumptions (i) and (ii) means that resonant modes can be
expressed as CW- or CCW- propagating wave modes along a ring trajectory
with identical resonant frequencies(i.e., degenerate states) 
in non-rotating cavities.

Recently, advanced semiconductor technologies make it possible to fabricate
arbitrary-sized optical cavities \cite{Yamamoto,Optical processes}.
When the cavity size becomes smaller, 
however, it remains unclear whether assumptions (i) and (ii)
are still valid and conventional theory based on them
really works.
In addition, resonant modes cannot always be
expressed as counter-propagating waves in microcavities.



We have recently established a wave-dynamical approach to
 rotating resonant cavities without assumptions (i) and (ii).
We showed that,  
when the angular velocity is larger than a certain value, 
the nearly-degenerate standing wave modes of a non-rotating microcavity
change into a pair of counter-propagating wave modes and
their frequency difference starts to increase proportionally to the angular 
velocity \cite{SH1}.
It has been revealed that, below the value of the angular velocity, 
a region exists where the frequency difference almost does not change, 
suggesting that the existence of this region might be an obstacle 
in the practical sense when one operates the microcavities 
as optical gyroscopes.


In this paper, we discuss the relation between the existence of this
region and the symmetry of cavity shape 
and show that designing cavity symmetry  
as $C_{nv}$ $(n\ge 3)$ yields degenerate resonant frequencies 
and can 
eliminate the region.

This paper is organized as follows: 
In Sec. \ref{sec:II}, we review the derivation
for the shifts of resonant frequencies induced by rotation and the
numerical results by a
wave-dynamical approach without assumptions (i) and (ii).
It is shown that there exists a region around zero angular velocity  
where the frequency shifts are almost not caused.
Discussions are provided in Sec. \ref{sec:III}, and we propose a method for
eliminating the region in Sec. \ref{sec:IV}. 
Finally, a summary of this paper is provided in Sec. \ref{sec:V}. 
\section{Theory \label{sec:II}}
According to the electromagnetic equations of
naturally covariant forms \cite{Anderson} or general theory of
relativity \cite{Landau,Heer}, electromagnetic fields in a rotating
resonant microcavity are subject to the Maxwell equations generalized to
a non-inertial frame of reference in uniform rotation with angular
velocity vector
$\vec{\Omega}$. 
We derived the following stationary wave
equation from the Maxwell equations by assuming that microcavities 
can be modeled as a waveguide, which is wide in the $xy$ direction 
and thin in the $z$ direction \cite{SH1}:
\begin{eqnarray}
\left(
{\nabla_{xy}}^2 + n^2 k^2
\right)\psi
-
2ik
\left(\vec{h}\cdot\nabla
\right)\psi
= 0, 
\label{fundeq}
\end{eqnarray}
where $\vec{h}$ $= (\vec{r}\times\vec{\Omega})/c$.
$\vec{r}$ $=(x,y,z)$ denotes coordinates in
a rotating frame of reference with angular velocity vector 
$\vec{\Omega}=(0,0,\Omega)$. 
The cavity is at rest in the rotating frame of reference.
 $c$ and $n$ are respectively
the velocity of the light and a refractive index inside the cavity, and
$k$ is the wave number. We also assume that the TM or TE waves of the
electromagnetic fields oscillate as 
$\vec{E}(\vec{r},t) [\vec{H}(\vec{r},t)] 
= (0,0,\psi(\vec{r})\exp(-ickt)+c.c)$. 
Below, we impose the Dirichlet boundary condition for simplicity, and 
resonant modes of rotating cavities can be obtained 
by solving Eq. (\ref{fundeq}) with two perturbation theories typically used 
in quantum mechanics.

First, we discuss a case where angular velocity $\Omega$ is
 smaller than the following value $\Omega_{th}$:
\begin{equation}
 \Omega_{th} 
=
n^2
\left| 
\int \int_D
\left[
\psi_1 
\left(
y\dfrac{\partial }{\partial x}-x\dfrac{\partial }{\partial y}
\right)
\psi_2 
\right] dxdy
\right|^{-1}c\Delta k_0, 
\label{ineq}
\end{equation}
where $D$ denotes 
the domain of the cavity, and 
$\Delta k_0$ is the spacing between adjacent eigenvalues $k$
of the wave number for Eq. (\ref{fundeq}) with $\Omega=0$,
i.e., 
$
\left(
{\nabla_{xy}}^2 + n^2 k^2
\right)\psi
=
0. 
$
 $\psi_1$ and $\psi_2$ are the wave functions 
of these eigenstates (resonant modes) that correspond to the adjacent
eigenvalues. In this case, we apply the perturbation theory for
non-degenerate states to Eq. (\ref{fundeq}) by assuming that, 
due to cavity rotation, the eigenvalue is shifted as
$k=k_1 +\delta k +O(|\Omega/c|^2)$ and 
the wave function is changed as 
$
\psi = \psi_1 + \delta \psi.
$
By substituting them to Eq. (\ref{fundeq}), we obtain
\begin{equation}
 \delta k= 
\dfrac{1}{n^2k_1}\int\int_{D}
\left[
\psi_1 \left(\vec{h}\cdot\nabla\right)\psi_1
\right]dxdy
=
0
\end{equation}
up to the first order of $|\Omega/c|$ 
and $\delta \psi = \sum_{i\ne 1}c_i\psi_i$, where
\begin{eqnarray}
c_i = 
\dfrac{2i {k_1} \int\int_{D}
\left[
\psi_i \left(\vec{h}\cdot\nabla\right)\psi_1
\right]dxdy
}
{ n^2({k_1}^2-{k_i}^2)}.
\label{cn}
\end{eqnarray}
These results mean that, as long as angular velocity $\Omega$ is small
enough,  
frequency shift due to rotation does not occur, 
$\delta \omega = c\delta k =0$, 
and the wave functions
of the eigenstates do not change CW- or CCW- rotating
waves.

Next we discuss the case where angular velocity is larger 
than $\Omega_{th}$, i.e., $\Omega\gg\Omega_{th}$.
In this case, the perturbation theory used above breaks down because 
 coefficient $c_i$ of resonant mode $i=2$, which has the
adjacent eigenvalue to $k_1$, is no longer small
due to $|c_2| \approx |\Omega/\Omega_{th}| \gg 1$.
Instead,
we use the perturbation theory for degenerate states.
According to the perturbation theory, 
two nearly-degenerate wavefunctions $\psi_1$ and $\psi_2$ are superposed
to reproduce the solutions of Eq. (\ref{fundeq}) as follows:
\begin{eqnarray}
\psi_{\pm} = \dfrac{1}{\sqrt{2}}
\left(
\psi_1 \pm i\psi_2
\right),
\label{cw}
\end{eqnarray} 
where the wavenumbers $k$ are shifted by rotation:
\begin{eqnarray}
k_{\pm} =\frac{k_1+k_2}{2}\pm \dfrac{1}{n^2}
\left|\int\int_{D}
\left[
\psi_1 \left(\vec{h}\cdot\nabla\right)\psi_2 
\right]dxdy
\right|. \label{eq:k-sag}
\end{eqnarray}
Accordingly, the frequency difference 
between the two eigenstates newly produced by cavity rotation 
is proportional to the angular velocity \cite{Explain}: 
\begin{equation}
\Delta \omega = 2
\left|\int\int_{D}
\left[
\psi_1 \left(y\frac{\partial}{\partial x}
-x\frac{\partial}{\partial y}\right)\psi_2 
\right]dxdy
\right|
\frac{\Omega}{n^2}, \label{eq:w-sag}
\end{equation}
where $\Delta\omega = c|k_+-k_-|$.
%
%

\begin{figure}
\begin{center}
\raisebox{0.0cm}{\includegraphics[width=4cm]{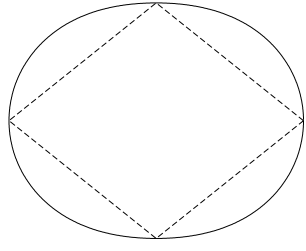}}
\end{center}
\caption{\label{fig-shring} 
(Solid line) Optical cavity defined by boundary 
$R(\theta) = R_0(1+\epsilon\cos 2\theta)$,
where $(r,\theta)$ denotes cylindrical coordinates.
The parameters of the cavity are set as $R_0=6.2866\mu m$, 
$\epsilon=0.12$, and $n=1$.
Dashed line denotes a ring trajectory.
}
\end{figure}

\begin{figure}
\begin{center}
  \begin{tabular}{ c c }
\raisebox{0.0cm}{\includegraphics[width=4.0cm]{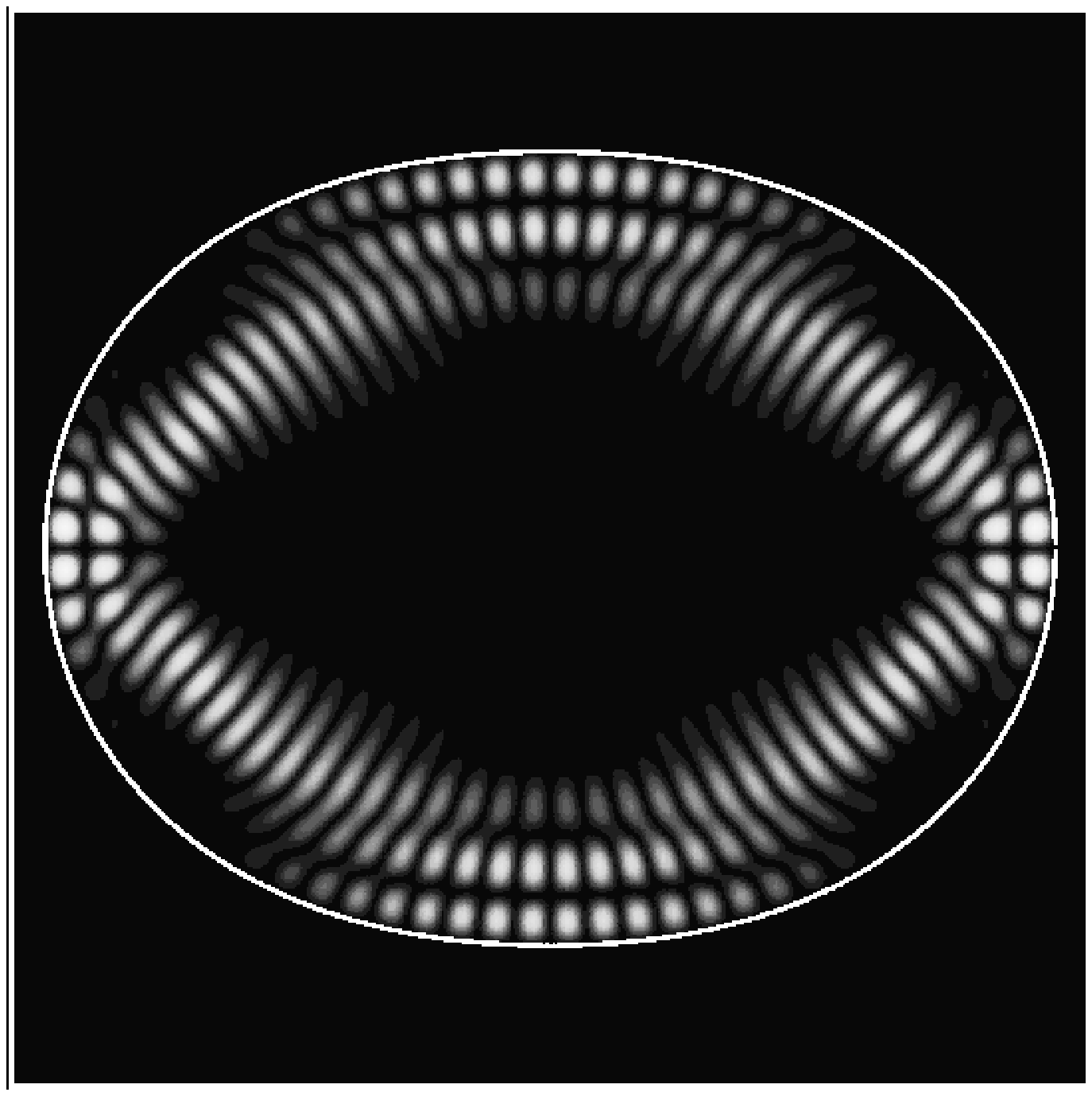}}
    &
    \includegraphics[width=4.0cm]{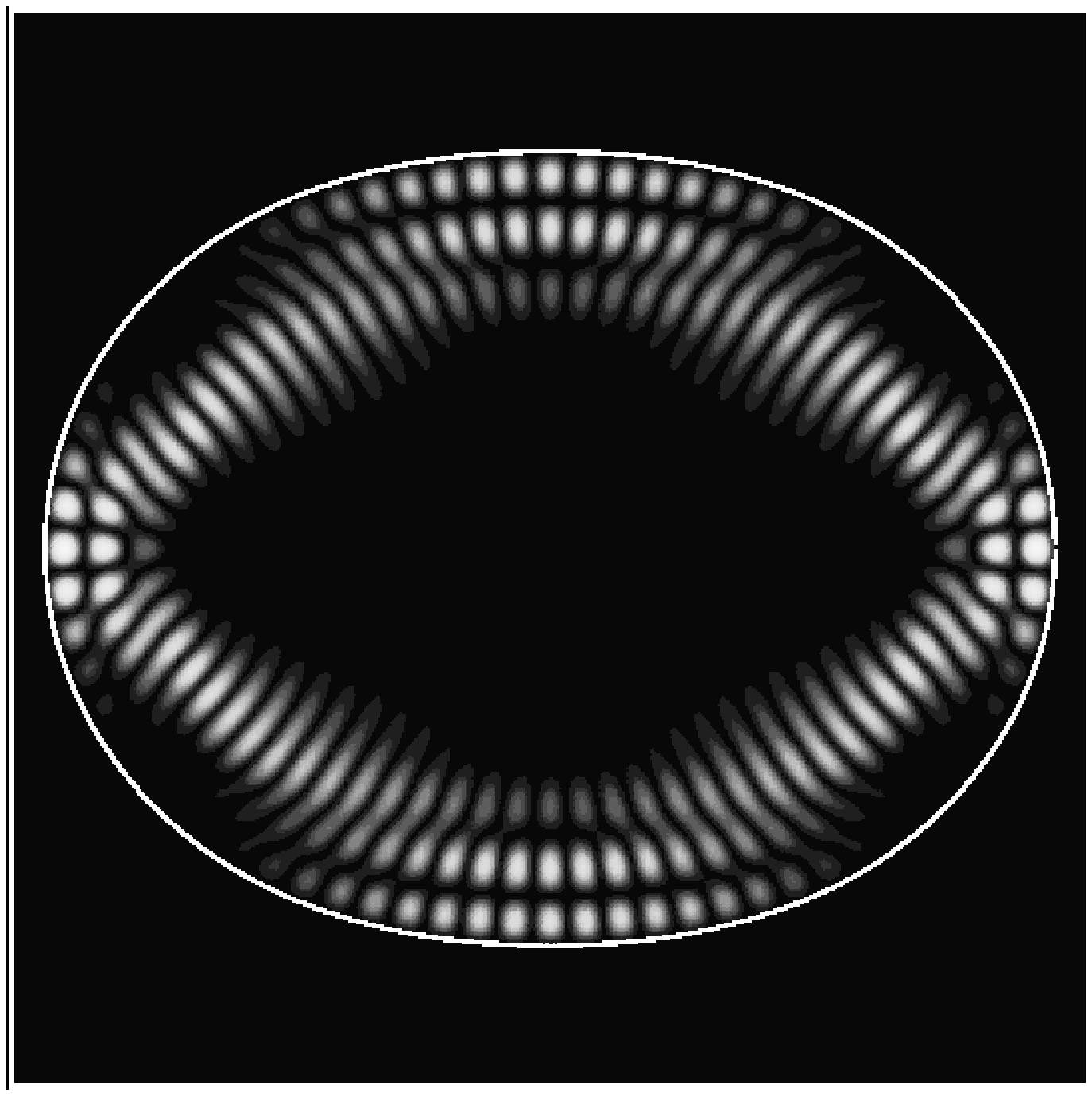}\\
    (a) & (b) 
  \end{tabular}
\end{center}
\vspace{-4mm}
\caption{\label{fig1} 
Wave functions of eigenstates localized around ring trajectory
 shown in Fig. \ref{fig-shring}.
(a) (Dimensionless) eigen-wavenumber $nk_AR_0$ is $49.3380585$ and 
the wave function has odd (odd) parity with respect to horizontal
 (vertical) axis.
(b) Eigen-wavenumber $nk_BR_0=49.3380615$ and even (even) 
parity with respect to horizontal (vertical) axis. 
White curves denote cavity boundary.}
\end{figure}

To demonstrate our theoretical results, we
carry out the numerical simulation in a
quadrupole cavity defined by boundary  
$R(\theta) = R_0(1 + \epsilon\cos 2\theta)$, 
where $(R,\theta)$ denotes the cylindrical coordinates. 
The parameters of the cavity are set as $\epsilon=0.12$ and the
refractive index is $n=1$,
where there is a ring trajectory shown by the
dashed line in Fig. \ref{fig-shring}. 
When the cavity is not rotated, 
solving Eq. (\ref{fundeq}) with $\Omega=0$ yields
the wavefunctions of the eigen modes corresponding to the ring
trajectory, as shown in Figs. \ref{fig1}(a) and (b).
As seen here, these modes are standing waves along the ring
trajectory and have different parities with respect to the (horizontal
and vertical) symmetry axes of the cavity. In addition, they have slightly
different eigenvalues of the wavenumber; these modes are
nearly-degenerate standing waves. We call the two modes shown in
Figs. \ref{fig1} (a) and (b) modes A and B, respectively. 

\begin{figure}
\begin{center}
\raisebox{0.0cm}{\includegraphics[width=9cm]{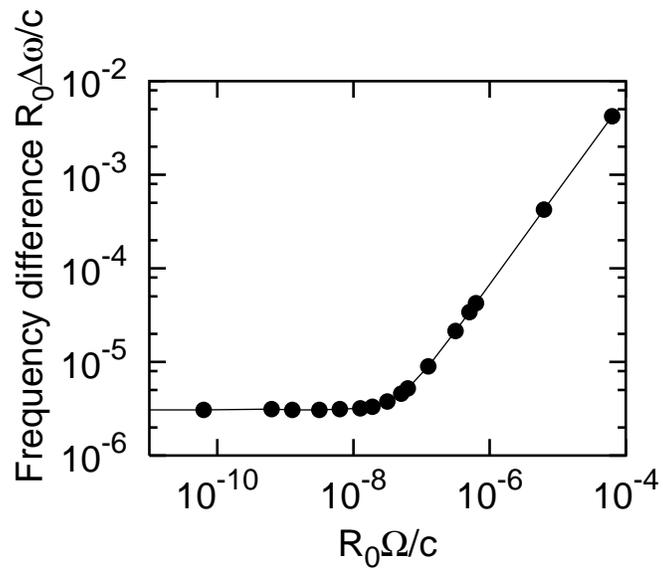}} \\
\end{center}
\caption{\label{fig4}
(a) (Dimensionless) frequency difference 
$R_0\Delta\omega/c$ 
versus 
 (dimensionless) angular velocity $R_0\Omega/c$.
Frequency difference does not change for 
$R_0\Omega/c < R_0\Omega_{th}/c (\sim 5\times 10^{-8})$.
For $R_0\Omega/c >R_0\Omega_{th}/c$, 
it becomes proportional to angular velocity $\Omega$.
}
\end{figure}

\begin{figure}
\begin{center}
  \begin{tabular}{ c c }
 \raisebox{0.0cm}{\includegraphics[width=4.0cm]{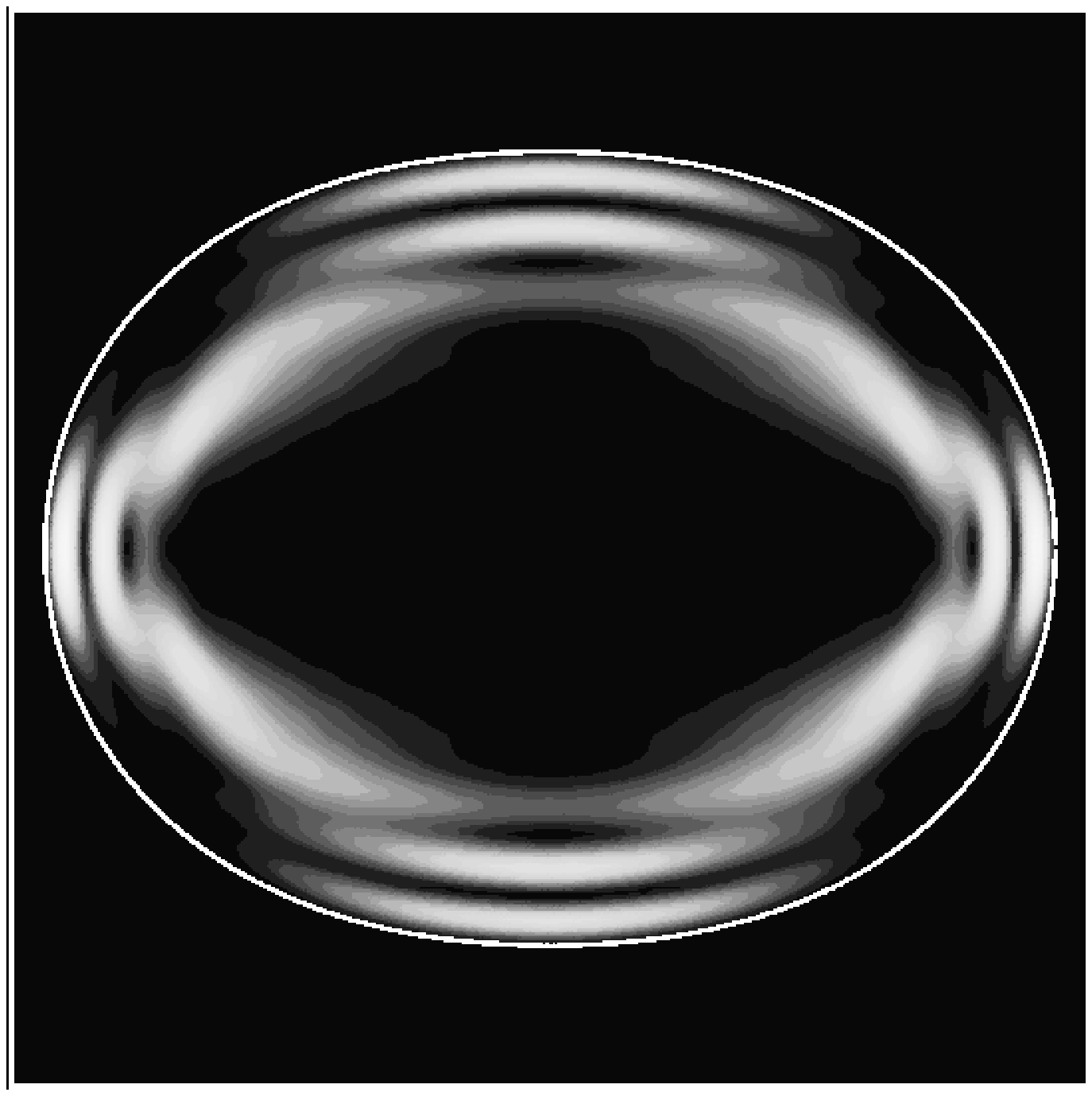}}
    &
    \includegraphics[width=4.0cm]{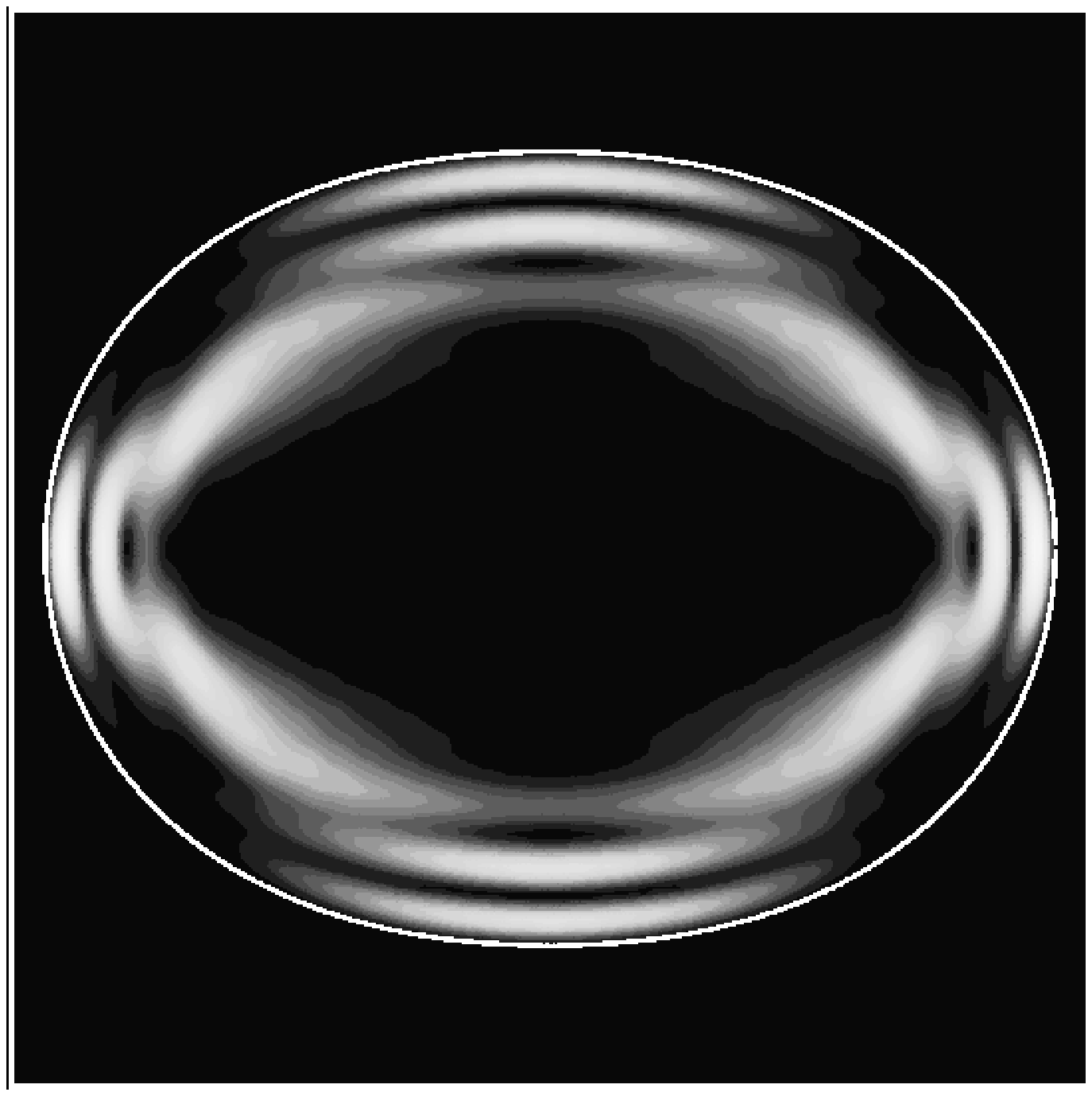}\\
    (a) & (b) 
  \end{tabular}
\end{center}
\caption{\label{fig3} 
Wave functions of rotating cavity 
with (dimensionless) angular velocity  
$R_0\Omega/c\approx 6.28\times 10^{-5} (> R_0\Omega_{th}/c)$
respectively corresponding to (a) mode A and (b) mode B.
}
\end{figure}

When the cavity is rotated but the angular velocity $\Omega$ is smaller
than $\Omega_{th}$ (where $R_0\Omega_{th}/c\sim 5\times 10^{-8}$), 
the frequency difference between modes A and B 
does not increase, as shown in Fig. \ref{fig4}(a)
, and the wavefunctions of modes A and B
do not drastically change, and they remain standing waves. 
However, for
$\Omega>\Omega_{th}$, the frequency difference increases proportionally
to the angular velocity $\Omega$. Then the wavefunctions of modes A and B are
superimposed following Eq. (\ref{cw}) and change
into the wavefunctions shown in Figs. \ref{fig3}(a) and (b),
respectively. The wavefunctions corresponds to the CW- and CCW-
rotating waves \cite{Explain2}. 
From these results, one can observe the Sagnac effect (in cases of resonant
cavities) as the frequency difference between counter-propagating wave modes
$\Delta \omega$ proportional to the angular velocity. 
Note that a region of angular
velocity $\Omega\ll \Omega_{th}$ exists 
where the frequency difference almost does not change.
This result can never be derived from the conventional theory of the
Sagnac effect using assumptions (i) and (ii).

\section{Discussion \label{sec:III}}
Note that the existence of region $\Omega<\Omega_{th}$
 is not akin to 
the so-called ``lock-in'' phenomenon that
 has been frequently observed in ring laser gyroscopes
\cite{Crow,Aronowitz,Lamb2,Spreeuw2}.
The lock-in phenomenon in ring laser gyros 
is that CW- (CCW-) rotating wave modes are
injection-locked to CCW-(CW-) propagating wave modes by backscatters in
the ring lasers at a low rotation rate.
However, in our analysis, 
 the backscatters and the nonlinear effect of 
laser medium are not both taken into account.
Our results show that a region of $\Omega<\Omega_{th}$ shown in
 the previous section exists
even without backscatters and the nonlinearity of laser medium.

As seen in Eq. (\ref{ineq}), 
 the existence of region $\Omega<\Omega_{th}$ 
originates in the existence of non-degeneracy (or
spacing $\Delta k_0$).
It is well-known that spacing $\Delta k_0$ 
is determined by the ratio between 
the characteristic radius of cavity $R_0$ 
and wavelength $\lambda$ of the light, $R_0/\lambda$ (for example,
Ref. \cite{Hackenbroich}). 
Angular velocity $\Omega_{th}$ increases as cavity size
decreases.   
In such a case, the existence of the $\Omega <\Omega_{th}$ region
might prevent the optical cavity 
from operating as an optical gyroscope, because the frequency difference
 almost does not change for angular velocity $\Omega$ 
smaller than $\Omega_{th}$. 

Recently, in a semiconductor microcavity laser called the quasi-stadium laser, 
the frequency difference between nearly-degenerate ring modes 
has experimentally measured as the frequency of oscillatory behavior,
which the intensities of the CW- and CCW- output lights periodically 
oscillate with
$\pi$ phase difference \cite{Muhan}. 
The frequency difference 
($=c\Delta k_0/(2\pi)$) was around 3 $MHz$, although the value
was changed by the pumping power. Applying Eq. (\ref{ineq}) to the case
of the quasi-stadium cavity, $\Omega_{th}$ can be
estimated as the order of $10^{6} (degree/second)$, 
which is so large that one might not be able to experimentally 
observe frequency shift due to rotation even with any equipments.
In the next section, we propose a method to remove the 
$\Omega<\Omega_{th}$ region.

\section{Application: design of resonant microcavity gyroscopes 
\label{sec:IV}}
According to Eq. (\ref{ineq}),  
one can see that $\Omega_{th}$ equals zero in the case of 
\begin{equation}
\Delta k_0 = 0. \label{eq8}
\end{equation}
The simplest case that satisfies condition (\ref{eq8}) is when
a cavity has continuous symmetry, such as circular symmetry, 
(e.g., micro ring or microdisk) because  
there are degenerate CW- and CCW- rotating waves
in the non-rotating cavity.
%
%
In a real system, however, 
the cavity must be coupled with 
an apparatus, such as a coupler or a photo detector,
to actually measure beat frequency (frequency difference) 
between the counter-propagating waves.  
The total configuration of the cavity and such
an apparatus cannot have continuous symmetry.
Therefore, the break in symmetry causes frequency splitting even in the absence of
rotation, i.e., $\Delta \omega_0 (=c\Delta k_0)\ne 0$. 
Thus, coupling the cavity with continuous symmetry with an
apparatus complicates satisfying condition (\ref{eq8}).
Accordingly, to preserve the existence of degeneracy
even in cases of total configuration,  
the cavity must have at most discrete symmetry
because 
it is possible to preserve the discrete symmetry of 
the configuration by symmetrically deposing the apparatus.

\subsection{Degeneracy and irreducible representations of symmetry group
  of cavity}

The existence of degenerate eigenstates in a cavity with discrete
symmetry can be shown by applying the representation theory of groups to wave
Eq. (\ref{fundeq}) with $\Omega=0$ \cite{group1}.
Here we are concerned with the finite point group
that leaves the cavity invariant in two dimensions.

 An irreducible representation (IR)
of the symmetry group of the cavity can 
characterize an eigenstate of Eq. (\ref{fundeq}) with $\Omega=0$ \cite{Sakanaka}.
Then the degree of degeneracy of
the eigenvalue equals the dimension of the IR. 
That is,  when the symmetry group of the cavity has a two-dimensional IR,
the wave equation can give a degenerate eigenstate corresponding to
the IR. 

Based on group theory, the point groups (rotations and reflections) 
in two dimensions can have one-dimensional and two-dimensional
IR's. 
A requirement for the existence of a two-dimensional IR is that 
the group is not commutative.
If the symmetry group of the cavity is commutative
group, such as $C_i$ symmetry, $C_n$ ($n=1,2,\cdots$) symmetry 
or $C_{nv}$ $(n\le 2)$ symmetry,
the group only has one-dimensional IR's, namely,
no degenerate eigenstates (barring accidental degeneracy). 
Note that the symmetry of the cavity shown in Fig. \ref{fig-shring}
is classified into $C_{2v}$. 
Since there are only non-degenerate eigenstates in the cavity,
the $\Omega<\Omega_{th}$ region is caused by the lack of degeneracy, as
shown in the previous section.
Thus, the only symmetry group that is not commutative is
 $C_{nv}$ ($n\ge 3$). 
%
%

Here, let $R_n$ and $\sigma$ be the rotation operator of $2\pi/n$ around
the origin of the coordinate system and the reflection symmetry operator
with respect to a symmetry axis, respectively.
Then, we denote wavefunctions that have even and odd parities
with respect to the symmetry axis by $\psi_+$ and $\psi_-$,
respectively, as 
\begin{eqnarray}
\sigma\psi_{\pm} = \pm\psi_{\pm} \label{eq:parities}.
\end{eqnarray} 
%
Note that even and odd parity wavefunctions $\psi_{+}$ and $\psi_{-}$ 
are always the solutions of Eq. (\ref{fundeq}) 
with $\Omega=0$ in a cavity with a symmetry axis. 
In addition, 
 also note that wavefunctions $\psi_+$ and $\psi_-$ are orthogonal. 
In a cavity with $C_{nv}$ ($n\ge 3$) symmetry,
the eigenvalues of wavefunctions $\psi_{\pm}$, which satisfy the
following condition for integer $k \in \{ 1,2,\cdots n \}$, are
degenerate:
\begin{eqnarray}
R_n^{k}\psi_{\pm}
\ne
R_n^{-k}\psi_{\pm} \label{eq10}.
\end{eqnarray}
The reason is because $\psi_{-(+)}$ can be converted to $\psi_{+(-)}$ 
as follows: 
First, 
odd parity wavefunction $\psi_-$ is rotated by $(2\pi/n)k$ clockwise
and counterclockwise around the origin of the coordinate system. 
Rotated wavefunction $R_n^{\pm k} \psi_{-}$ also 
has the same eigenvalue as wavefunction $\psi_-$, 
because the cavity is invariant for rotational operators
$R_n^{\pm k}$.
Then linear combination $R_n^k\psi_--R_n^{-k}\psi_-$ also has the same 
eigenvalue.  Operating $\sigma$ to the linear combination yields the
following result:
\begin{eqnarray}
\sigma
(
R_n^k\psi_- - R_n^{-k}\psi_-
)
&=&
R_n^{-k}\sigma\psi_- - R_n^{k}\sigma\psi_- \nonumber \\
&=&
+(
R_n^k\psi_- - R_n^{-k}\psi_-
),
\end{eqnarray}
where $\sigma\psi_-=-\psi_-$ and formula $(\sigma R_n^k)^2=1$ are used.
This means that the linear combination is 
even parity with respect to the same symmetry axis. That is, 
even-parity wavefunction $\psi_+$ can be written as 
$\psi_+ =  c_k(R_n^k\psi_--R_n^{-k}\psi_-)$, where $c_k$ is a constant.
Even-parity wavefunction $\psi_+$ can also be converted to odd
parity wavefunction $\psi_-$ in the same way.
Accordingly,
one can see that condition (\ref{eq8}) can be satisfied 
for the eigenstates of the wavefunctions 
satisfying condition (\ref{eq10}) in the cavity with $C_{nv}$ ($n\ge 3$) 
symmetry and that $\Omega_{th}$ becomes zero in this case.

\subsection{Numerical simulation}
\begin{figure}
\begin{center}
\raisebox{0.0cm}{\includegraphics[width=8cm]{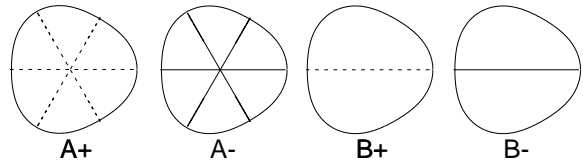}}
\end{center}
\vspace{-4mm}
\caption{\label{fig:parity} Four parity symmetry classes of 
 a $C_{3v}$ symmetric cavity defined by the boundary 
$R(\theta) = R_0(1+\epsilon\cos 3\theta)$, where $\epsilon=0.065$. 
Even (odd) symmetry is marked by dashed (solid) lines.
}
\end{figure}

\begin{figure}
\begin{center}
  \begin{tabular}{ c c }
\raisebox{0.0cm}{\includegraphics[width=4cm]{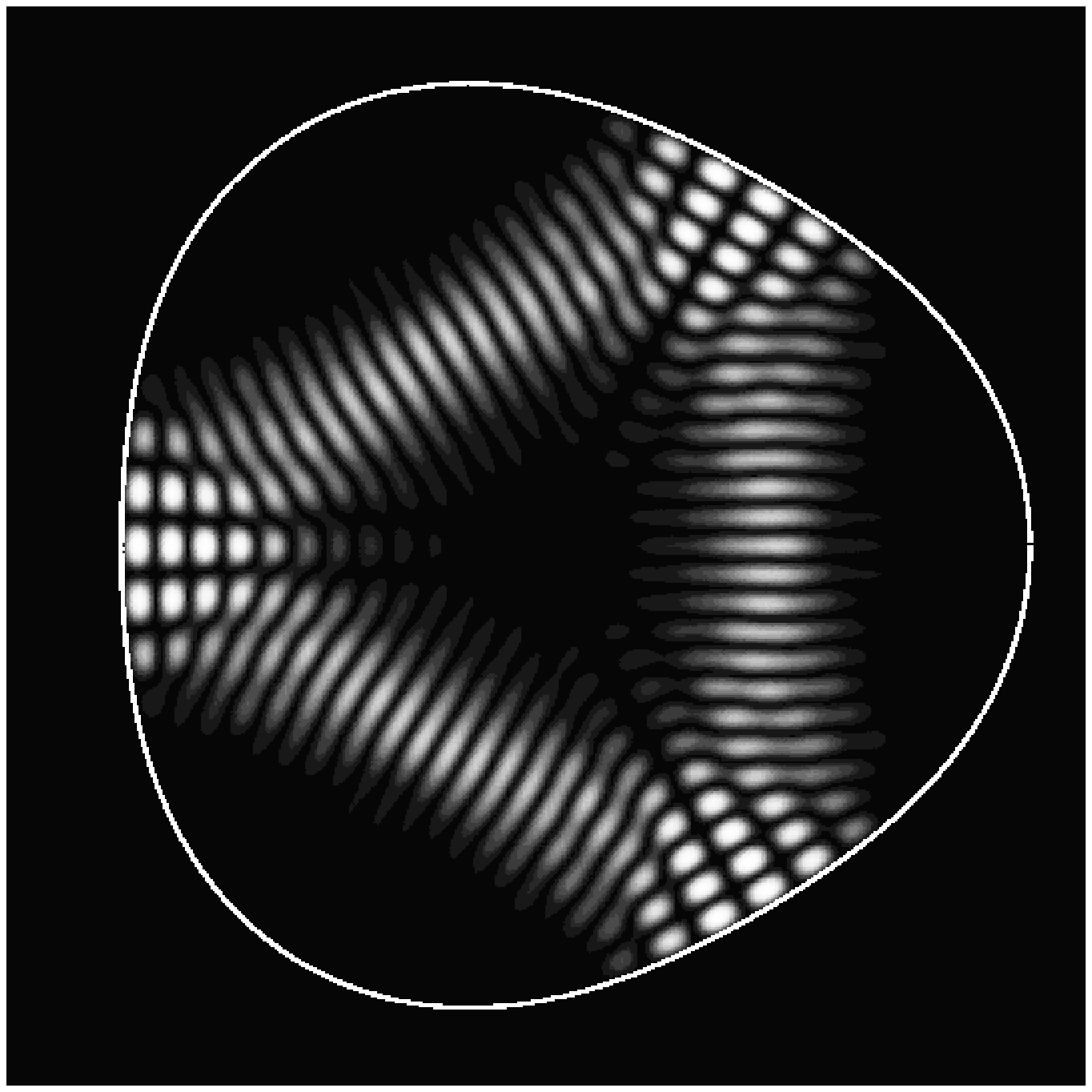}}
    &
    \includegraphics[width=4cm]{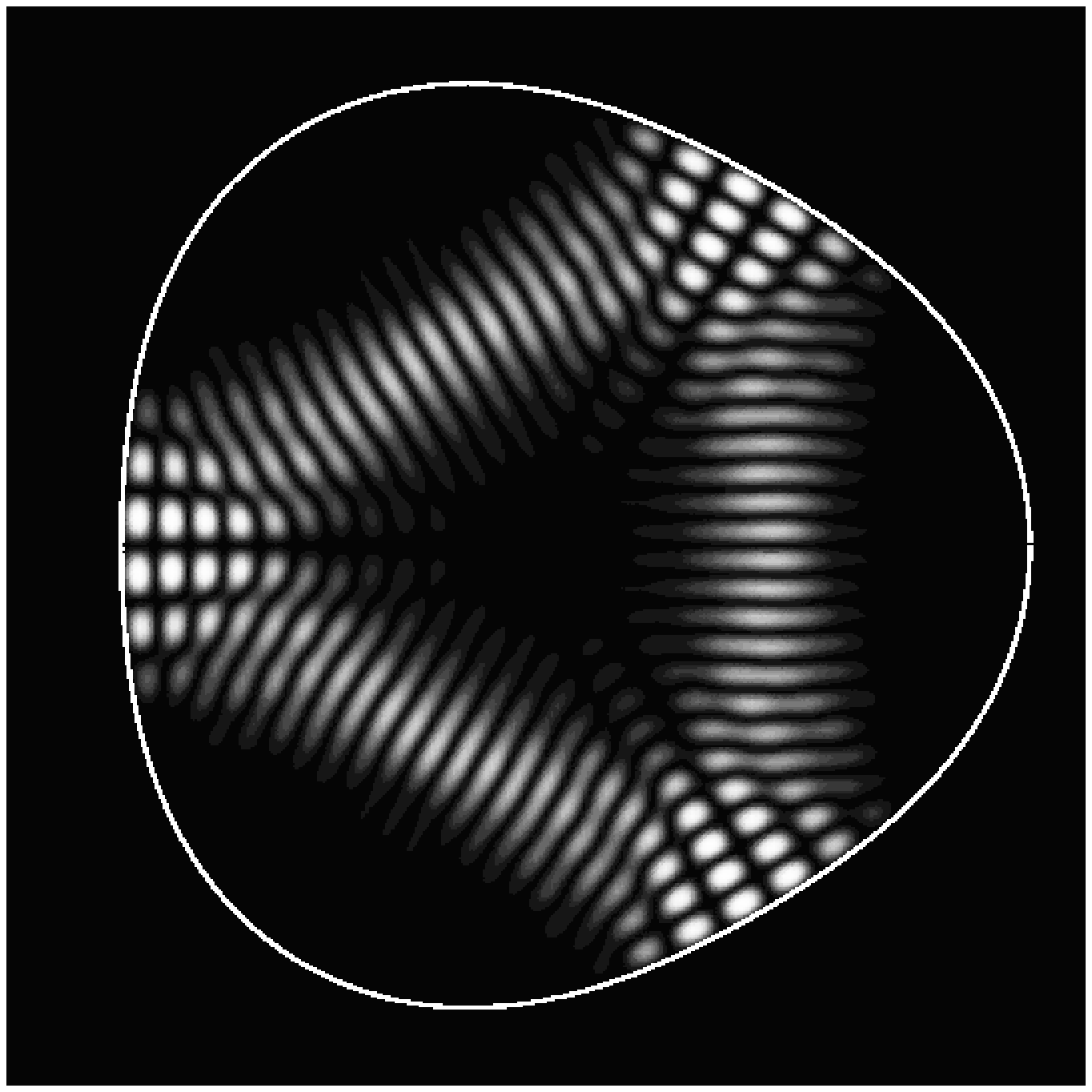}\\
    (a) & (b) \\
\raisebox{0.0cm}{\includegraphics[width=4cm]{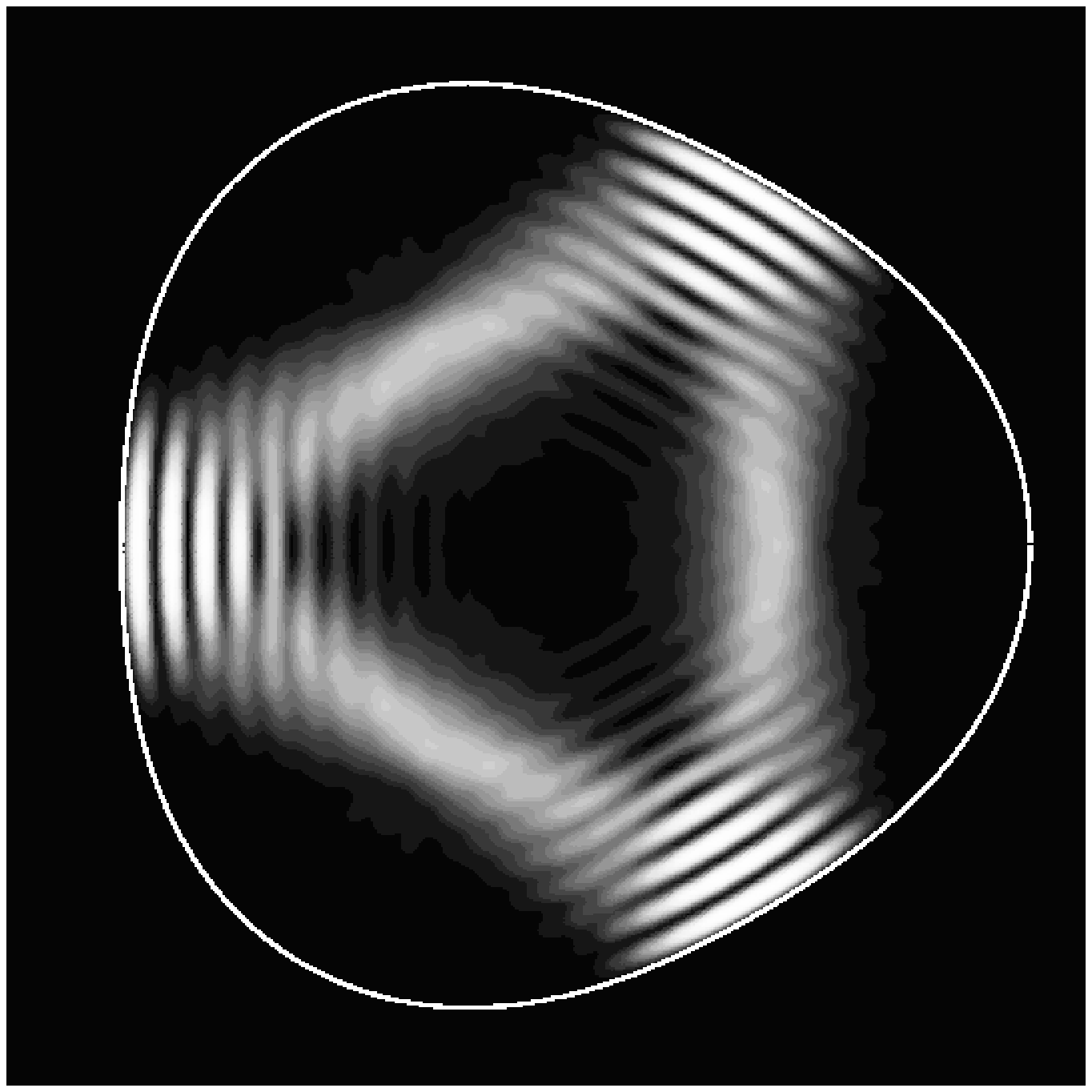}}
    &
    \includegraphics[width=4cm]{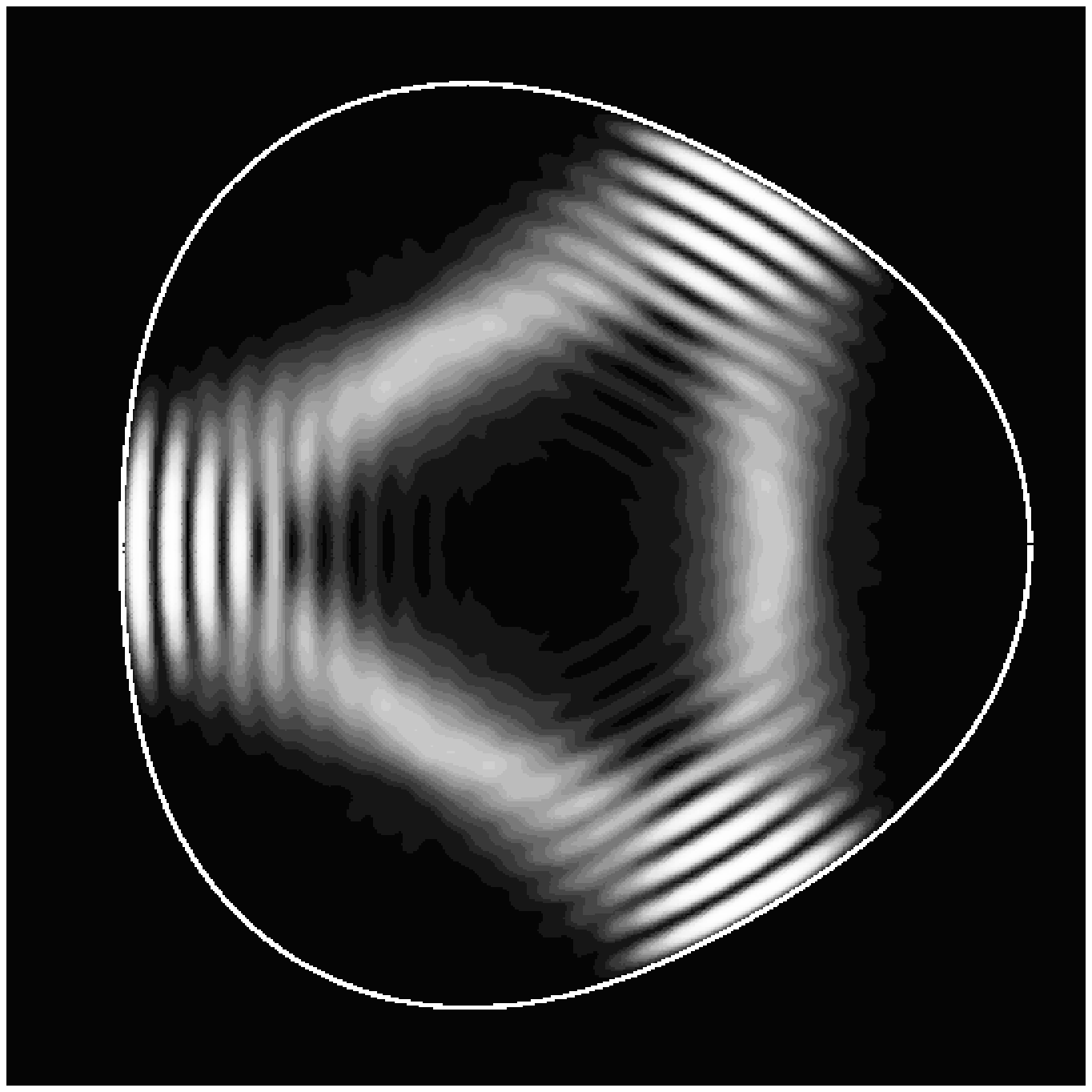}\\
    (c) & (d) \\

  \end{tabular}
\end{center}
\vspace{-4mm}
\caption{\label{C3vwave} (a-b) Degenerate standing-wavefunctions
 of (dimensionless) eigen-wavenumber $nkR_0=50.220063$ in non-rotating cavity. 
(c-d) Wavefunctions of rotating cavity
 with $R_0\Omega/c\approx 6.28\times 10^{-11}$. 
White curves denote cavity boundary.}

\end{figure}

Theoretical predictions are confirmed by numerical simulation in which 
we chose a cavity defined by boundary
$R(\theta)=R_0(1+\epsilon \cos 3\theta)$
that has $C_{3v}$ symmetry, where $\epsilon =0.065$.  
Figure \ref{fig:parity} shows the four parity symmetry classes of this
cavity.
In notation $A+,\cdots$, the sign is $+(-)$ if the wave
function is even (odd) with respect to the horizontal axis.
Letter $A$ indicates a wavefunction that does not satisfy 
condition (\ref{eq10}), while letter $B$ indicates a wave function that does.
Therefore, solving Eq. (\ref{fundeq}) with $\Omega=0$ 
yields two degenerate standing-wave eigenfunctions, as shown in
Figs. \ref{C3vwave}(a) and (b), which are
classified into $B+$ and $B-$, respectively.  

\begin{figure}
\begin{center}
\raisebox{0.0cm}{\includegraphics[width=9cm]{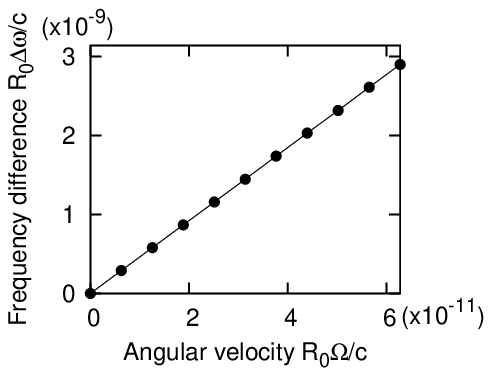}}
\end{center}
\vspace{-4mm}
\caption{\label{C3vfreq}  (Dimensionless) frequency difference
 $R_0\Delta\omega/c$ v.s.  (dimensionless) angular velocity $R_0\Omega/c$.
}
\end{figure}

For $\Omega>0$, standing-wave 
eigenfunctions, as shown in Figs. \ref{C3vwave}(a) and (b),
change into the rotating wavefunctions shown
in Figs. \ref{C3vwave}(c) and (d), respectively.
Then, as shown in Fig.~\ref{C3vfreq}, 
the frequency difference between the two eigenstates 
is proportional to angular velocity $\Omega$.
Accordingly, the frequency difference can be observed
in a rotating cavity with $C_{3v}$ symmetry 
for $\Omega\ge \Omega_{th}=0$.

\section{Summary \label{sec:V}}
In summary, we showed that designing the symmetry of cavity shape as $C_{3v}$ 
$(n\ge 3)$ symmetry gives degenerate resonance frequencies in
non-rotating cavities and eliminates region 
$\Omega<\Omega_{th}$ 
where there is no resulting the shifts of resonant frequencies 
proportional to rotation rate.

Recently, semiconductor ring lasers \cite{Sorel1,Cao}
and various types of microcavities \cite{Randy,Armenise,Steinberg1}
 have been studied experimentally and theoretically toward the
 realization of small-sized rotation velocity sensors. 
We believe that our results will be useful for designing 
such rotation velocity sensors.

\noindent {\bf Acknowledgments}\\
We thank T. Miyasaka for discussions and numerical simulations.
The work was supported by 
the National Institute of information and Communication Technology
of Japan. 


\begin{thebibliography}{99}
\bibitem{Explain}
Equation (\ref{eq:w-sag}) agrees with a conventional expression for 
frequency difference: $(4A\omega)/(cnP)\Omega$ by applying 
assumptions (i) and (ii) to Eq. (\ref{eq:w-sag}). 
In the above, $A$ is the area bounded by the optical path, 
$P$ is the perimeter of the optical path, 
and $\omega$ is the eigen frequency of the non-rotating ring cavity 
approximated as $\omega \approx c(k_1+k_2)/2$.
\end{thebibliography}
\end{document}